\documentclass[10pt]{article}

\usepackage{amssymb,amsbsy}
\usepackage{amsmath}
\usepackage{amsthm}
\usepackage{graphics,graphicx}
\usepackage[T1]{fontenc}
\usepackage{color}

\newcommand{\beq}{\begin{equation}}
\newcommand{\eeq}{\end{equation}}

\newcommand{\beqar}{\begin{eqnarray}}
\newcommand{\eeqar}{\end{eqnarray}}
\newcommand{\bit}{\begin{itemize}}
\newcommand{\eit}{\end{itemize}}
\newcommand{\benum}{\begin{enumerate}}
\newcommand{\eenum}{\end{enumerate}}
\newcommand{\barr}{\begin{array}}
\newcommand{\earr}{\end{array}}
\def\ds{\displaystyle}

\newcommand{\jump}[2]{[\mbox{\hspace{-#1em}}[#2]\mbox{\hspace{-#1em}}]}

\newcommand{\bjump}[2]{\left[\mbox{\hspace{-#1em}}\left[#2\right]\mbox{\hspace{-#1em}}\right]}

\newcommand{\modIII}{\text{III}}

\def\XXint#1#2#3{{\setbox0=\hbox{$#1{#2#3}{\int}$}
   \vcenter{\hbox{$#2#3$}}\kern-.5\wd0}}

\def\b0{\mbox{\boldmath $0$}}

\def\be{\mbox{\boldmath $e$}}

\def\bn{\mbox{\boldmath $n$}}

\def\bp{\mbox{\boldmath $p$}}
\def\bq{\mbox{\boldmath $q$}}

\def\bt{\mbox{\boldmath $t$}}
\def\bu{\mbox{\boldmath $u$}}

\def\bI{\mbox{\boldmath $I$}}

\newcommand{\bsigma}{\mbox{\boldmath $\sigma$}}

\newcommand{\bepsilon}{\mbox{\boldmath $\epsilon$}}
\newcommand{\bvarepsilon}{\mbox{\boldmath $\varepsilon$}}
\newcommand{\btau}{\mbox{\boldmath $\tau$}}

\newcommand{\bvarphi}{\mbox{\boldmath$\varphi$}}

\newcommand{\bchi}{\mbox{\boldmath $\chi$}}
\newcommand{\bmu}{\mbox{\boldmath $\mu$}}

\def\f0{\ensuremath{\mathbb{O}}}

\newcommand{\tr}{\mathop{\mathrm{tr}}}

\newcommand{\diver}{\mathop{\mathrm{div}}}

\newcommand{\curl}{\mathop{\mathrm{curl}}}

\def\EFM{{\it Eng.\ Fract.\ Mech.}\ }

\def\IJF{{\it Int.\ J.\ Fracture}\ }

\def\IJSS{{\it Int.\ J.\ Solids Struct.}\ }

\def\JAM{{\it ASME J.\ Appl.\ Mech.}\ }

\def\JMPS{{\it J.\ Mech.\ Phys.\ Solids}\ }

\title{Mode III interfacial crack in the presence of couple-stress elastic materials}

\author{A. Piccolroaz$^{(1)}$, G. Mishuris$^{(1)}$, E. Radi$^{(2)}$
\\
\\
$^{(1)}$
{\it Institute of Mathematical and Physical Sciences, Aberystwyth University,} \\ 
{\it Ceredigion SY23 3BZ, Wales, U.K.} \\
$^{(2)}$
{\it Dipartimento di Scienze e Metodi dell'Ingegneria, Universit\`a di Modena e Reggio Emilia } \\
{\it Viale Amendola 2, I-42122 Reggio Emilia, Italia} \\
}

\begin{document}

\maketitle

\begin{abstract}
\noindent
In this paper we are concerned with the problem of a crack lying at the interface between dissimilar materials 
with microstructure undergoing antiplane deformations. The micropolar behaviour of the materials is described by 
the theory of couple-stress elasticity developed by Koiter (1964). This constitutive model includes the 
characteristic lengths in bending and torsion and thus it is able to account for the underlying microstructure of 
the two materials. We perform an asymptotic analysis to investigate the behaviour of the solution near the crack 
tip. It turns out that the stress singularity at the crack tip is strongly influenced by the microstructural 
parameters and it may or may not show oscillatory behaviour depending on the ratio between the characteristic 
lengths. 
\end{abstract}


\newpage

\tableofcontents
 
\newpage

{\bf Nomenclature}

\vspace{10mm}
\begin{tabular}{ll}
$G_\pm$ & shear modulus ($+/-$ stands for upper/lower half-plane) \\
$\nu_\pm$ & Poisson's ratio \\
$l_b^\pm$ & material characteristic length in bending \\
$l_t^\pm$ & material characteristic length in torsion \\
$l_\pm$ & material characteristic length (Koiter's notation)\\
$\eta_\pm$ & ratio between the characteristic lengths in bending and torsion (Koiter's notation, $-1 < \eta_\pm \leq 1$)  \\
$\bu$ & displacement field \\
$w$ & out-of-plane displacement \\
$\bvarphi$ & rotation vector \\
$\bepsilon$ & strain tensor \\
$\bchi$ & curvature tensor \\
$\bt$ & nonsymmetric stress tensor \\
$\bsigma$ & symmetric part of the stress tensor \\
$\btau$ & skew-symmetric part of the stress tensor \\
$\bmu$ & couple-stress tensor \\
$\bp$ & reduced tractions vector \\
$\bq$ & reduced couple-stress tractions vector \\
$\lambda$ & order of stress singularity
\end{tabular}

\newpage

\section{Introduction}
\label{sec1}

Nowadays, bimaterials are efficiently and widely used in many advanced engineering applications, such as layered 
composite structures, electronic packaging and thin film coatings. For the prediction of failure of these 
structures and the assessment of acceptable stress level under the condition experienced during service, it 
becomes essential to estimate the magnitude and distribution of the interfacial stress and strain fields along the 
interface and mainly near the tip of interface cracks, which may arise and extend under general loading 
conditions. In particular, antiplane shear loading condition may frequently occur in the life span of composite 
structures, both alone or accompanied by plane deformation. 

Within the classical LEFM theory, the crack tip fields for an interface crack under antiplane strain are similar 
to the Mode III crack tip fields in a homogeneous medium (Willis, 1971; Piccolroaz et al., 2009). In both cases, 
indeed, the shear stresses on the crack plane are the same in the upper and lower bodies, whereas the out-of-plane 
displacement is zero on the uncracked region of the crack plane. Thus, it is possible to combine the lower and the 
upper bodies to obtain equilibrium, without changing displacements or stresses in the two halves. Stresses exhibit 
Mode III symmetry, but displacements do not and thus crack sliding profiles are not symmetric. 

Due to the lack of a length scale, the classical theory of elasticity is not able to characterize the constitutive 
behaviour of brittle materials at the micron scale. This lack is expected to be particularly significant for the 
analysis of the stress and deformation fields very near the crack tip. For a proper investigations of the 
crack tip fields at the micron scale it becomes necessary to adopt enhanced constitutive models, which 
account for the presence of microstructure. A way of doing that consists in the inclusion of one or more 
characteristic lengths, typically of the same order of the compositional grain size, generally few microns, for 
many advanced materials. The indeterminate theory of couple-stress elasticity (CSE) developed by Koiter (1964) 
involves the material characteristic lengths in bending and torsion. It is sufficiently accurate to simulate the 
behaviour of materials at the micron scale as well as the size effects occurring at distances to the crack tip 
comparable to characteristic lengths, but it is also simple enough to allow the achievement of closed-form 
solutions.

Although the presence of the microstructure is expected to modify the interface crack tip field with respect to 
the classical solution of the LEFM, no analytical investigations have been so far performed about the problem of 
an antiplane crack along the interface between micropolar and classical elastic materials (the only related work regards a 
crack terminating perpendicular to the interface; Mishuris, 1985). Most of related studies 
available in literature instead concern the problem of an interface crack under plane deformations, e.g., Itou 
(1991) examined the effect of couple-stresses on the strain energy release rate for an interface crack loaded by 
an internal pressure, neglecting somehow the oscillatory behaviour of the crack tip fields. 

In order to provide an experimental basis for studying the interfacial behaviour of a bimaterial specimen under 
shear loading, Kang et al.\ (2002) applied a method which combines moir\'e interferometry with phase shift and 
image processing to measuring the interfacial displacement and strain fields within the interfacial region. Their 
experimental results show that there is a boundary layer characteristic with a peak value of shear strain and high 
gradient of rotation angle in the interfacial region. Their study also shows that similar results can be 
analytically predicted by means of the couple-stress theory, considering the additional freedom of the rotation 
angle effect. 

Hutapea et al.\ (2003) investigated the micro-stress generated along a fiber/matrix interface under generalized 
plane deformation, which are expected to dominate the failure initiation process in composite laminate. They 
showed that the micropolar theory is able to capture the interface micro-stress accurately.

A small number of interface crack problems have been investigated by using the strain gradient theory of 
plasticity (Hao and Liu, 1999; Chen and Wang, 2002). In particular, Hao and Liu (1999) analyzed the crack 
propagation in bimaterial systems showing that high stress triaxiality always occurs on the softer material, which 
may promote ductile damage and facilitate crack growth. Chen and Wang (2002) explored the interface crack tip 
fields at micron scales under plane strain conditions. Their numerical investigations show that the singularity of 
stresses in the strain gradient theory slightly exceeds or equals to the square-root singularity independently of 
the material hardening exponents. Askes and Gitman (2009) showed numerically that in gradient 
elasticity no singular behaviour is found for a crack terminating perpendicular to the interface.

The analysis of singular stress concentration in homogeneous micropolar elastic solids shows that several 
pathological predictions of classical elasticity in singular stress concentration problems are altered, mitigated, 
or possibly eliminated when couple-stresses are taken into account (Nazarov and Semenov, 1980). 
In particular, the problem of a Mode III crack 
in a homogeneous materials modelled by the couple-stress elastic theory was first analyzed by Zhang et al. (1998) 
and later by Geogiadis (2003) by considering a single characteristic length. The results obtained therein indicate 
that the skew-symmetric stress components have $r^{-3/2}$ singularity near the crack tip, where $r$ is the 
distance to the crack tip. Although this singularity is much stronger than the conventional square-root 
singularity, it does not violate the boundness of strain energy surrounding the crack tip and leads to a finite 
energy release rate. Their asymptotic analysis also provides a negative out-of-plane displacement ahead of the 
crack tip. This unphysical result is due to the exclusion of the lowest order terms for the displacement and 
symmetric stress components, which do not contribute to the energy release rate.

The effects of both characteristic lengths in bending and torsion and a complete investigation of the crack tip 
fields under Mode III loading condition in homogeneous CSE materials have been properly addressed in a recent work 
by Radi (2008). The roles of both characteristic lengths are therein examined in detail and their influence on the 
crack tip is analytically explored by using Fourier transform and the Wiener-Hopf method. The asymptotic and 
full-field analyses show that the symmetric stress is finite at the crack tip, whereas the skew-symmetric stress 
is negative and strongly singular. Ahead of the crack tip within a zone smaller than the characteristic length in 
torsion, both the total shear stress and reduced tractions occur with the opposite sign with respect to the 
classical LEFM solution, as predicted by the asymptotic analysis. However, the zone of dominance of the asymptotic 
fields has limited physical relevance and becomes vanishing small for a characteristic length in torsion of zero. 
In this limit, the full-field solution recovers the classical $K_\modIII$ field with square-root stress 
singularity. Outside this zone, the total shear stress exhibits a positive maximum, thus providing more realistic 
predictions on the tractions level ahead of the crack tip than the singular LEFM solution. A sharp crack profile 
is also observed. It may denote that the crack becomes stiffer, thus revealing that the presence of 
microstructures may shield the crack tip from fracture.

In the present work, the effects of strain rotation gradients on a stationary antiplane crack along the interface 
between two different couple-stress elastic materials are analytically investigated by performing an asymptotic 
analysis of the crack-tip fields. The special problem of a crack along the interface between a couple-stress 
elastic solid and a classical elastic medium is also addressed in Sec.\ \ref{sec3}. The results of the present 
asymptotic analysis are expected to hold in a small zone near to the crack tip, whose extent may vary with the 
size of the characteristic lengths, and provide valuable information for performing a full-field analysis of the 
interface crack problem, e.g., by using the Wiener--Hopf method, which will be the subject of further 
investigations.

\section{Crack at the interface between couple-stress elastic materials}
\label{sec2}

\begin{figure}[!htcb]
\begin{center}
\includegraphics[width=8cm]{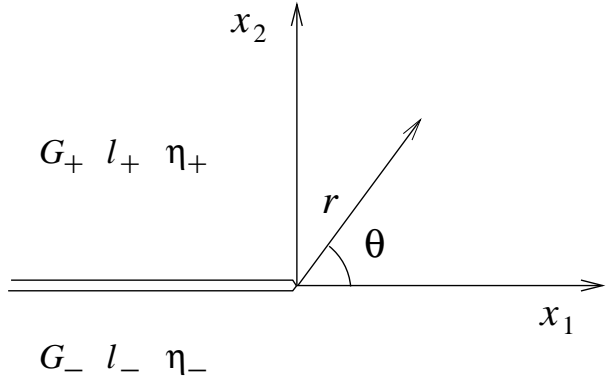}
\caption{\footnotesize A crack at the interface between dissimilar couple-stress materials.}
\label{fig01}
\end{center}
\end{figure}

We consider a bimaterial plane made of two dissimilar materials, joined along a perfect interface. The two 
materials are assumed to have an underlying microstructure, described by the material characteristic lengths in 
bending and in torsion, denoted by $l_b^\pm$ and $l_t^\pm$, respectively. The elastic moduli are denoted by 
$G_\pm$ (shear modulus) and $\nu_\pm$ (Poisson's ratio). A semi-infinite plane crack is placed along the 
interface, and a Cartesian reference system is assumed centred at the crack tip, see Fig. \ref{fig01}.

The fundamentals of the Couple Stress (CS) elasticity theory (Koiter, 1964) can be found in several text books and 
research papers (see for example Nowacki, 1985; Asaro and Lubarda, 2006; Radi, 2008). It is recalled here that the 
main characteristic of this theory is that the rotation vector $\bvarphi$ is not independent of the displacement 
vector $\bu$, but it is subject to the condition 
\beq
\label{constraint}
\bvarphi = \frac{1}{2} \curl \bu.
\eeq 
Consequently, all the kinematical quantities can be derived from the displacement field. In particular, for 
antiplane shear deformations, the following kinematical relations between the out-of-plane displacement $w$, 
rotation vector $\bvarphi$, strain tensor $\bvarepsilon$ and curvature tensor $\bchi$ are derived
\beq
\barr{l}
\ds
\varphi_1 = \frac{1}{2} \frac{\partial w}{\partial x_2}, \quad 
\varphi_2 = -\frac{1}{2} \frac{\partial w}{\partial x_1}, \quad
\varepsilon_{13} = \frac{1}{2} \frac{\partial w}{\partial x_1}, \quad
\varepsilon_{23} = \frac{1}{2} \frac{\partial w}{\partial x_2}, \\[5mm]
\ds
\chi_{11} = - \chi_{22} = \frac{1}{2} \frac{\partial^2 w}{\partial x_1 \partial x_2}, \quad
\chi_{21} = -\frac{1}{2} \frac{\partial^2 w}{\partial x_1^2}, \quad
\chi_{12} = \frac{1}{2} \frac{\partial^2 w}{\partial x_2^2}.
\earr
\eeq

An infinitesimal surface element transmits a force and a couple vector, which give rise to a nonsymmetric stress 
tensor, $\bt$, and a couple-stress tensor, $\bmu$. The nonsymmetric stress tensor can be decomposed into a 
symmetric part $\bsigma$ and a skew-symmetric part $\btau$, such that $\bt = \bsigma + \btau$. 

The isotropic constitutive equations are given by
\beq
\bsigma = \frac{2\nu G (\tr\bvarepsilon)}{1 - 2\nu} \bI + 2G \bvarepsilon, \quad 
\bmu = 2Gl^2 (\bchi^T + \eta \bchi),
\eeq
where $\bI$ is the identity tensor, $l$ and $\eta$ the CS parameters 
introduced by Koiter (1964), with $-1 < \eta \leq 1$, the superscript $^T$ denotes transposition. The material parameters $l$ and $\eta$ characterize the 
microstructure of the material and can be expressed in terms of the material characteristic lengths in bending 
and in torsion as follows:
\beq
l_b = l/\sqrt{2}, \quad l_t = l \sqrt{1 + \eta}.
\eeq
For antiplane shear deformations, the nonzero stress and couple-stress components are
\beq
\label{symstress}
\sigma_{13} = G \frac{\partial w}{\partial x_1}, \quad
\sigma_{23} = G \frac{\partial w}{\partial x_2}, 
\eeq
\beq
\label{couple}
\mu_{11} = -\mu_{22} = Gl^2 \left(1 + \eta\right) 
\frac{\partial^2 w}{\partial x_1 \partial x_2}, \quad
\mu_{12} = Gl^2 \left(\eta\frac{\partial^2 w}{\partial x_2^2} - 
\frac{\partial^2 w}{\partial x_1^2}\right), \quad
\mu_{21} = Gl^2 \left(\frac{\partial^2 w}{\partial x_2^2} - 
\eta\frac{\partial^2 w}{\partial x_1^2}\right).
\eeq
In the absence of body forces and body couples, the equations of equilibrium read
\beq
\label{equil}
\diver \bt^T = 0, \quad
\diver \bmu^T + (\btau\be_1) \times \be_1 + (\btau\be_2) \times \be_2 + 
(\btau\be_3) \times \be_3 = 0,
\eeq
where $\{\be_1,\be_2,\be_3\}$ is an orthonormal basis. For antiplane shear deformations, the nonzero 
skew-symmetric stress components, derived from (\ref{couple}) and (\ref{equil})$_2$, are
\beq
\label{skewstress}
\tau_{13} = -\frac{Gl^2}{2} \Delta \frac{\partial w}{\partial x_1}, \quad
\tau_{23} = -\frac{Gl^2}{2} \Delta \frac{\partial w}{\partial x_2},
\eeq
where $\Delta$ stands for the laplacian operator.

In the CS theory, due to the internal constraint (\ref{constraint}) between rotations and displacements, the 
Neumann boundary conditions are prescribed in terms of the so called reduced force tractions vector $\bp$ and 
couple tractions vector $\bq$ defined as 
\beq
\bp = \bt^T \bn + \frac{1}{2} \nabla \mu_{nn} \times \bn, \quad
\bq = \bmu^T \bn - \mu_{nn} \bn,
\eeq
respectively, where $\bn$ denotes the outward unit normal and $\mu_{nn} = \bn \cdot \bmu \bn$. Additionally, if the 
external surface of the body is not smooth but piecewise smooth, the boundary conditions include the following equation 
along each edge 
\beq
\label{edge}
Q = \frac{1}{2}(\mu_{nn}^+ - \mu_{nn}^-),
\eeq
where $Q$ is a prescribed line load tangential to the edge (Koiter, 1964), and superscripts $+$ and $-$ stand for the values 
on the surface at each side of the edge. It then appears that the condition (\ref{edge}) becomes essential in the  case of 
bodies with non-regular boundaries, such as cusps, wedges and cracks.

A substitution of (\ref{symstress}) and (\ref{skewstress}) into (\ref{equil})$_1$ gives the following governing 
equation for the displacements $w^\pm$ in the two half-planes:
\beq
\label{gov}
\Delta w^\pm - \frac{l_\pm^2}{2} \Delta^2 w^\pm = 0,
\eeq
where $\Delta^2$ denotes the bilaplacian operator.

We assume that the crack faces are traction-free, so that the following boundary conditions apply for 
$x_2 = 0^\pm$ and $x_1 < 0$:
\beq
\label{free}
\barr{l}
\ds
p_3^\pm := 
G_\pm \left\{ \frac{\partial w^\pm}{\partial x_2} - \frac{l_\pm^2}{2} 
\frac{\partial}{\partial x_2}\left[ (2 + \eta_\pm) \frac{\partial^2 w^\pm}{\partial x_1^2} 
+ \frac{\partial^2 w^\pm}{\partial x_2^2} \right] \right\} = 0, \\[5mm]
\ds
q_1^\pm := 
G_\pm l_\pm^2 \left\{ \frac{\partial^2 w^\pm}{\partial x_2^2} 
- \eta_\pm \frac{\partial^2 w^\pm}{\partial x_1^2} \right\} = 0.
\earr
\eeq
Assuming also that no tangential line load $Q$ is applied along the crack edge, we enforce that
\beq
\label{edge2}
\mu_{22}^+ - \mu_{22}^- = -G_+l_+^2(1 + \eta_+)\left.\frac{\partial^2 w^+}{\partial x_1 \partial x_2}\right|_{(x_1,x_2) = (0^-,0^+)} +
G_-l_-^2(1 + \eta_-)\left.\frac{\partial^2 w^-}{\partial x_1 \partial x_2}\right|_{(x_1,x_2) = (0^-,0^-)} = 0.
\eeq

The formulation is completed by the transmission conditions for ideal interface, which imply continuity of the 
displacements, rotations, reduced stress and couple-stress components for $x_2 = 0$ and $x_1 > 0$:
\beq
\label{tra}
\barr{l}
\ds
\jump{0.15}{w} = 0, \quad
\bjump{0.55}{\frac{\partial w}{\partial x_2}} = 0, \\[5mm]
\ds
\bjump{0.55}{G \left\{ \frac{\partial w}{\partial x_2} - \frac{l^2}{2} 
\frac{\partial}{\partial x_2}\left[ (2 + \eta) \frac{\partial^2 w}{\partial x_1^2} 
+ \frac{\partial^2 w}{\partial x_2^2} \right] \right\}} = 0, \quad 
\bjump{0.55}{G l^2 \left\{ \frac{\partial^2 w}{\partial x_2^2} 
- \eta \frac{\partial^2 w}{\partial x_1^2} \right\}} = 0,
\earr
\eeq
where the notation $\jump{0.15}{f}$ stands for the jump of the function $f$ across the interface: 
$\jump{0.15}{f} = f^+ - f^-$.

\subsection{Asymptotic analysis and singularity of stresses}
\label{sec21}

Assuming a polar reference system centered at the crack tip, we search for the main asymptotic term of the 
solution as $r \to 0$ in the standard form as follows
\beq
\label{main}
w^\pm(r,\theta) = r^\lambda F_\pm(\theta,\lambda).
\eeq

We are interested in finding the leading term of the asymptotic solution corresponding to finite elastic energy. 
This requires that $\lambda \geq 3/2$ (see Radi, 2008). It is noted that the values $\lambda = 0$ and 
$\lambda = 1$ are also admissible, as long as the respective terms in (\ref{main}) correspond to a rigid body 
motion (constant displacement) and a uniform deformation (linear displacement), respectively. Moreover, the 
expression (\ref{main}) can be used to find more terms in the asymptotic solution in the form 
$\sum_i r^{\lambda_i} F_\pm(\theta,\lambda_i)$, provided that $|\lambda_i - \lambda_j| < 2$, $\forall\ i \neq j$. 
If more terms are required with exponents differing by 2 or more than 2, then a two-terms asymptotic procedure 
should be used instead, as explained in Sec.\ \ref{sec3}.

Keeping into account only the leading term as $r \to 0$, the governing equation (\ref{gov}) yields the following 
ODE for the unknown functions $F_\pm$
\beq
\label{ode}
F_\pm'''' + 2(\lambda^2 - 2\lambda + 2)F_\pm'' + \lambda^2(\lambda - 2)^2 F_\pm = 0.
\eeq
We first investigate the simplest cases $\lambda = 0$ and $\lambda = 1$, for which eq.\ (\ref{ode}) admits the 
solutions
\beq
\label{zero}
F_\pm(\theta,0) = B_1^\pm + B_2^\pm \theta + B_3^\pm \sin 2\theta + B_4^\pm \cos 2\theta, 
\eeq
\beq
\label{uno}
F_\pm(\theta,1) = (B_1^\pm + B_2^\pm \theta)\sin\theta + (B_3^\pm + B_4^\pm \theta)\cos\theta,
\eeq 
respectively. Taking into account all boundary and transmission conditions, one can conclude that for 
$\lambda = 0$ and $\lambda = 1$, eqs.\ (\ref{zero}) and (\ref{uno}) take, as expected, the forms
\beq
F_\pm(\theta,0) = \beta_0, \quad
F_\pm(\theta,1) = \beta_{11} \sin \theta + \beta_{12} \cos \theta,
\eeq
respectively. 

One can also use the representation (\ref{main}) to find the solution for the case $\lambda = 2$, since the term 
corresponding to $\lambda = 0$ is not involved in the analysis (as it vanishes after differentiation). Thus, for 
$\lambda = 2$, eq.\ (\ref{ode}) admits the solution
\beq
F_\pm(\theta,2) = B_1^\pm + B_2^\pm \theta + B_3^\pm \sin 2\theta + B_4^\pm \cos 2\theta,
\eeq
which, taking into account all boundary and transmission conditions, reduces to
\beq
F_\pm(\theta,2) = \beta_{21} (\cos^2\theta + \eta_\pm \sin^2\theta) + \beta_{22} \sin 2\theta.
\eeq
For all other cases, eq.\ (\ref{ode}) admits the following solution:
\beq
F_\pm(\theta) = B^\pm_1 \sin(\lambda\theta) + B^\pm_2 \cos(\lambda\theta) + 
B^\pm_3 \sin[(\lambda-2)\theta] + B^\pm_4 \cos[(\lambda-2)\theta].
\eeq
By imposing the boundary and transmission conditions, we obtain a 8$\times$8 homogeneous algebraic system, whose 
characteristic equation is
\beq
\label{char}
\sin^2(\pi \lambda) \left[ \cos(2\pi \lambda) + \kappa \right] = 0,
\eeq
where $\kappa = C/D > 0$ for any $\eta_\pm > -1$ and 
\beq
\barr{l}
\ds
\barr{l}
C = G_+^2l_+^4(5 - 2\eta_- + \eta_-^2)(3 - \eta_+)^2(1 + \eta_+)^2 + 
G_-^2l_-^4(5 - 2\eta_+ + \eta_+^2)(3 - \eta_-)^2(1 + \eta_-)^2 \\[2mm] 
+ 2G_+l_+^2G_-l_-^2(3 + \eta_+ + \eta_- - \eta_+\eta_-)(3 - \eta_+)(1 + \eta_+)
(3 - \eta_-)(1 + \eta_-),
\earr \\[7mm]
\ds
\barr{l}
D = (3 - \eta_+)(1 + \eta_+)(3 - \eta_-)(1 + \eta_-)
\left\{G_+^2l_+^4(3 - \eta_+)(1 + \eta_+) + G_-^2l_-^4(3 - \eta_-)(1 + \eta_-) \right. \\[2mm] 
\left. + 2G_+l_+^2G_-l_-^2(5 - \eta_+ -\eta_- + \eta_+\eta_-)\right\}.
\earr
\earr
\eeq

The first term in eq.\ (\ref{char}) leads to the conclusion that $\lambda = 3$ (the cases $\lambda = 0,1,2$ have 
been investigated above), while the second term may exhibit singular behaviour depending on the value of the 
parameter $\kappa$. If $\kappa > 1$, then the solution of eq. (\ref{char}) is complex and the singularity shows 
oscillatory behaviour in the vicinity of the crack tip. Otherwise, the solution is real and there are no 
oscillations. More precisely, since $\kappa$ is strictly positive, the following three cases may occur:
\begin{itemize}
\item[(i)] $0 < \kappa < 1$: the first admissible value of the exponent is in the interval $3/2 < \lambda < 7/4$ (simple root).
\item[(ii)] $\kappa = 1$: the first admissible value of the exponent is $\lambda = 3/2$ (double root).
\item[(iii)] $\kappa > 1$: the first admissible value of the exponent is $\lambda = 3/2 \pm i \gamma$ (simple root), where 
\beq
\gamma = \frac{1}{2\pi} \log(\kappa + \sqrt{\kappa^2 -1}).
\eeq
\end{itemize}

In the case of a homogeneous material, $G_+ = G_-$, $l_+ = l_-$, $\eta_+ = \eta_-$, the ratio $\kappa$ is equal to 
1 and thus the first admissible value for the exponent is 3/2 (this case has been analysed in Radi, 2008). Some 
other special cases are investigated in the next section.

\subsection{Some particular and special cases}
\label{sec22}

To decrease number of parameters, let us first consider the case where $\eta_+ = \eta_- = \eta$. Then the ratio 
$\kappa$ reduces to 
\beq
\label{kappa2}
\kappa = \frac{(G_+^2l_+^4 + G_-^2l_-^4)(5 - 2\eta + \eta^2) + 
2 G_+l_+^2 G_-l_-^2 (3 - \eta)(1 + \eta)}
{(G_+^2l_+^4 + G_-^2l_-^4)(3 - \eta)(1 + \eta) + 
2 G_+l_+^2 G_-l_-^2 (5 - 2\eta + \eta^2)}
= \frac{(a^2 + b^2)c + 2ab}{a^2 + b^2 + 2abc},
\eeq
where we use the notations $a = G_+l_+^2$, $b = G_-l_-^2$ and 
\beq
c(\eta) = \frac{5 - 2\eta + \eta^2}{(3 - \eta)(1 + \eta)}.
\eeq

\begin{figure}[!htcb]
\begin{center}
\includegraphics[width=8cm]{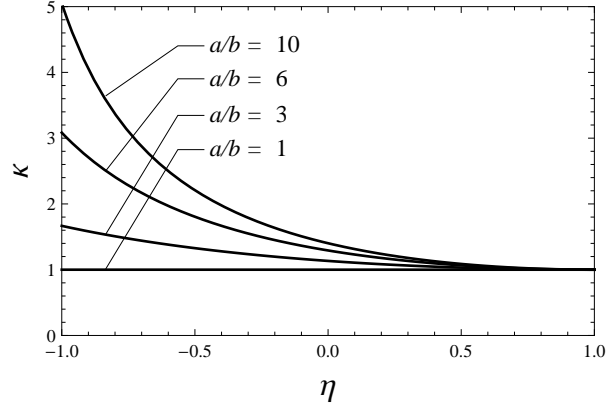}
\caption{\footnotesize Plot of the ratio $\kappa$ given by (\ref{kappa2}) as a function of $\eta$ for 
$a/b = 1,3,6,10$.}
\label{fig02b}
\end{center}
\end{figure}

Since $c \geq 1$ for any admissible value of $\eta$ ($-1 < \eta \leq 1$), it is easy to show that, for dissimilar 
materials, $\kappa$ is always greater than 1 and equal to 1 if and only if  $c = 1$ (or equivalently $\eta = 1$), 
see Fig.\ \ref{fig02b}.

In the limiting case $\eta = 1$ the asymptotic solution in the vicinity of the crack tip is given by
\beq
\label{solution}
\barr{l}
\ds w^\pm(r,\theta) = \beta_0 + r (\beta_1\sin\theta + \beta_2\cos\theta) + 
r^{3/2} \frac{l_\mp^2}{G_\pm} \left\{ \beta_3\left(3\sin\frac{\theta}{2} - 
\sin\frac{3\theta}{2}\right) \right. \\[3mm] 
\ds \left. + \beta_4 \left(\cos\frac{\theta}{2} - 
\cos\frac{3\theta}{2}\right) \right\} + r^2 (\beta_5 + \beta_6\sin 2\theta) + O(r^{5/2}), \quad r \to 0,
\earr
\eeq
where the constants $\beta_i$ are amplitude factors depending on far-field loading and specimen geometry. It is noted that 
logarithmic terms are excluded, since geometric and algebraic multiplicities of the double root coincide (R\"{o}ssle and S\"{a}ndig, 1996).

Correspondingly, the asymptotics of symmetric stress, couple-stress and skew-symmetric stress as $r \to 0$ are
\beq
\barr{l}
\ds
\sigma_{13}^\pm = \beta_2 G_\pm - 2 r^{1/2} l_\mp^2 \sin^2 \frac{\theta}{2} 
\left(3\beta_3\sin\frac{\theta}{2} + \beta_4\cos\frac{\theta}{2}\right) + 
2r G_\pm (\beta_5\cos\theta + \beta_6\sin\theta) + O(r^{3/2}), \\[5mm]
\ds
\sigma_{23}^\pm = \beta_1 G_\pm + r^{1/2} l_\mp^2 \sin \frac{\theta}{2} 
\left(3\beta_3\sin\theta + \beta_4(3 + \cos \theta)\right) + 
2r G_\pm (\beta_5\sin\theta + \beta_6\cos\theta) + O(r^{3/2}), \\[5mm]
\ds
\mu_{11}^\pm = - r^{-1/2} l_+^2 l_-^2 \sin \frac{\theta}{2} 
\left[3\beta_3(\sin\theta + \sin 2\theta) 
+ \beta_4(2 + \cos\theta + \cos 2\theta)\right] + 
4\beta_6 G_\pm l_\pm^2 + O(r^{1/2}), \\[5mm]
\ds
\mu_{21}^\pm = \mu_{12}^\pm = r^{-1/2} l_+^2 l_-^2  
\left\{\beta_3 \cos\frac{\theta}{2} (\sin 2\theta - \sin\theta) 
+ \beta_4 \cos\frac{\theta}{2} (2 - \cos\theta + \cos 2\theta)\right\} + O(r^{1/2}), \\[5mm]
\ds
\tau_{13}^\pm = \frac{1}{2} r^{-3/2} l_+^2 l_-^2 (3\beta_3\sin\frac{3\theta}{2} + 
\beta_4\cos\frac{3\theta}{2}) + O(r^{-1/2}), \\[5mm]
\ds
\tau_{23}^\pm = -\frac{1}{2} r^{-3/2} l_+^2 l_-^2 (3\beta_3\cos\frac{3\theta}{2} - 
\beta_4\sin\frac{3\theta}{2}) + O(r^{-1/2}).
\earr
\eeq
Applying now the condition (\ref{edge2}), we obtain
\beq
\label{edge3}
\mu_{22}^+(\theta = \pi) - \mu_{22}^-(\theta = -\pi) = \frac{4\beta_4 l_+^2 l_-^2}{\sqrt{r}} + 4\beta_6(G_-l_-^2 - G_+l_+^2) + 
O(r^{1/2}), \quad r \to 0.
\eeq
The limit, as $r \to 0$, of (\ref{edge3}) should equal the tangential line load $Q$ applied to the crack edge. It then appears that 
$\beta_4$ is always zero and $\beta_6$ does not vanish only if $Q$ is different from zero. It is also noted that the constant $\beta_3$ 
plays the role of stress intensity factor.

The asymptotics for the crack opening, $\jump{0.15}{w} = w^+(\theta=\pi) - w^-(\theta=-\pi)$, and for the 
skew-symmetric stress ahead of the crack tip, $\tau_{23}^\pm(\theta=0)$, as $r \to 0$, are given by
\beq
\jump{0.15}{w} = 4\beta_3 \frac{G_+l_+^2 + G_-l_-^2}{G_+G_-} r^{3/2} + O(r^{5/2}), \quad
\tau_{23}^\pm(\theta=0) = -\frac{3}{2} \beta_3 l_+^2 l_-^2 r^{-3/2} + O(r^{-1/2}). 
\eeq
Therefore, in the vicinity of the crack tip, the skew-symmetric stress at $\theta = 0$ displays a sign opposite to 
that of the crack opening, in contrast to the classical result of LEFM. A similar effect has been found by Radi 
(2008) in the case of a crack in an homogeneous CS material.

In the opposite case ($\eta_+ = \eta_- = \eta < 1$), the solution always exhibits oscillatory behaviour near the 
crack tip. Moreover, this behaviour is quite different from that we encounter in the case of classical materials. 
In the classic case, the region near the crack tip where the oscillatory behaviour appears is very small, while in 
the considered case this zone can be quite pronounced or its size can even tend to infinity if $\eta \to -1$. 
Taking into account that the asymptotic analysis given here is valid only in a small neighbourhood of the crack 
tip where the micropolar theory controls the behaviour of the solution, such situation has limited physical 
meaning. 

Let us now assume that the parameters $\eta_+$ and $\eta_-$ are different and one of them, say $\eta_-$, tends to 
the limiting value $\eta_- \to -1$, while the other is separated from $-1$, so that $\eta_+ > -1 + \epsilon$, 
where $\epsilon$ is a small positive parameter. In this case, one can easily check that $\kappa \to \infty$ for 
any fixed value of $\eta_+$. This again corresponds to the case where the exponent is a complex number with the 
imaginary part becoming infinite, $\gamma \to \infty$, and therefore it has no physical relevance.

It is noted that the case $\eta_- = -1$, $\eta_+ \neq -1$ cannot be recovered from the limiting case discussed 
above, and thus it will be discussed separately in Sec.\ \ref{sec33}.

Another case of interest is when one of the multipliers involved in the parameter $\kappa$, say $G_-l_-^2$, 
diminishes, $G_- l_-^2 \to 0$, then 
\beq
\label{uno2}
\kappa \to \frac{5 - 2\eta_- + \eta_-^2}{(3 - \eta_-)(1 + \eta_-)} = c(\eta_-) \geq 1.
\eeq
Once again, this solution has physical relevance if and only if $\eta_- = 1$. The assumption $G_-l_- \to 0$ takes 
place if one assumes that $l_- \to 0$, or, in other words, when the material in the lower half-plane reduces to a 
classical elastic material. This suggests considering the problem of a crack at the interface 
between a micropolar material (occupying the upper half-plane) and a classical one (occupying the lower half-
plane). However, it is not possible to recover from (\ref{solution}) the solution for a classical elastic 
material, due to the singular perturbation character of the equation (\ref{gov}) as $l_\pm \to 0$. 
For this reason, the problem of a crack lying at the interface between couple-stress elastic and classical elastic 
materials is addressed separately in the next section.

\subsection{Energy release rate}
\label{sec23}
 
\begin{figure}[!htcb]
\begin{center}
\includegraphics[width=8cm]{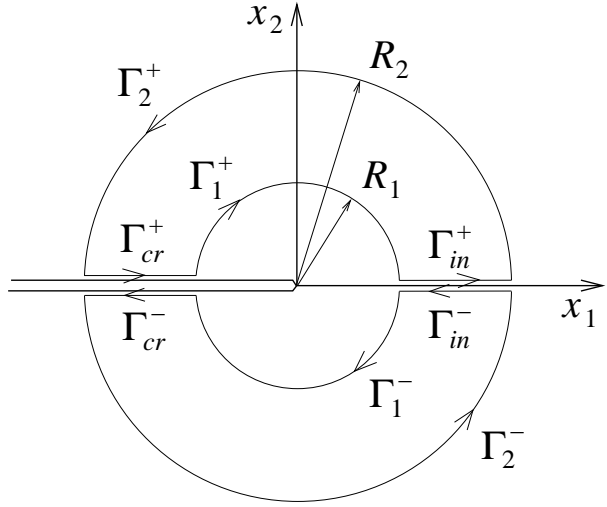}
\caption{\footnotesize Path of integration for the evaluation of the energy release rate.}
\label{fig03}
\end{center}
\end{figure}

In this section, the energy release rate is evaluated for the asymptotic representation (\ref{solution}), valid in
the case $\eta_+ = \eta_- = 1$, by means of the $J$-integral argument. The conservation law for couple-stress elasticity (see 
Lubarda and Markenscoff, 2000) implies that
\beq
\label{jint}
J = \int_\Gamma \left(W n_1 - \bt^T \bn \cdot \frac{\partial \bu}{\partial x_1} - 
\bmu^T \bn \cdot \frac{\partial \bvarphi}{\partial x_1}\right) ds = 0,
\eeq
for every closed contour $\Gamma$, provided that there is no singularity within $\Gamma$. In (\ref{jint}), $\bn$ 
is an outward unit normal on $\Gamma$ and $W$ denotes the strain-energy density
\beq
W = G \bepsilon \cdot \bepsilon + Gl^2 (\bchi \cdot \bchi + \eta \bchi \cdot \bchi^T).
\eeq
We define $\Gamma^\pm = \Gamma_1^\pm \cup \Gamma_2^\pm \cup \Gamma_{cr}^\pm \cup \Gamma_{in}^-$ (see Fig. \ref{fig03}), so that 
\beq
\label{cons}
\sum_\pm (J_2^\pm \mp J_{cr}^\pm - J_1^\pm \pm J_{in}^\pm) = 0, 
\eeq
with evident meaning of the symbols.
Then we evaluate each term according to the representation (\ref{solution}), thus obtaining
\beq
J_1 = J_1^+ + J_1^- = -\frac{2\beta_2\beta_4 l_+^2 l_-^2}{\sqrt{R_1}} + P + O(R_1^{1/2}),
\eeq
\beq
J_2 = J_2^+ + J_2^- = -\frac{2\beta_2\beta_4 l_+^2 l_-^2}{\sqrt{R_2}} + P + O(R_2^{1/2}),
\eeq
\beq
J_{cr}^\pm = \pm\frac{\beta_2\beta_4 l_+^2 l_-^2}{\sqrt{R_1}} \mp \frac{\beta_2\beta_4 l_+^2 l_-^2}{\sqrt{R_2}} 
+ O(R_2^{1/2}),
\eeq
\beq
J_{in}^\pm = -\frac{3\beta_2\beta_3 l_+^2 l_-^2}{\sqrt{R_1}} + \frac{3\beta_2\beta_3 l_+^2 l_-^2}{\sqrt{R_2}} 
+ O(R_2^{1/2}),
\eeq
in which
\beq
P = \sum_\pm \frac{l_\pm^2 l_\mp^4}{2G_\pm} (9\pi\beta_3^2 + \pi \beta_4^2 \pm 12 \beta_3\beta_4).
\eeq
Note that the integrals along the interface cancel out from (\ref{cons}), whereas the integrals along the crack faces give
\beq
J_{cr}^- - J_{cr}^+ = F(R_2) - F(R_1),
\eeq
where
\beq
F(R) = \frac{2\beta_2\beta_4 l_+^2 l_-^2}{\sqrt{R}} + O(R^{1/2}).
\eeq
It is possible now to introduce a path-independent parameter as follows:
\beq
J_* = \lim_{R_2 \to 0} (J_2 + F(R_2)) = \lim_{R_1 \to 0} (J_1 + F(R_1)) = P.
\eeq
In consideration of the condition (\ref{edge2}), $\beta_4$ vanishes and the path-independent parameter $J_*$ gives the energy release rate
(derived through direct energy balance considerations by Atkinson and Leppington, 1977 and Eshelby, 1980) 
\beq
J_* = \frac{9}{2} \pi l_+^2 l_-^2 \left(\frac{l_-^2}{G_+} + \frac{l_+^2}{G_-}\right) \beta_3^2.
\eeq

In the case of homogeneous body ($G_+ = G_- = G$, $l_+ = l_- = l$), our formula coincides with earlier results by Radi (2008):
\beq
J_* = \frac{9\pi l^6}{G} \beta_3^2.
\eeq

It is noted here that particular attention should be paid when using the $J$-integral in CS materials. The contribution of the integrals along 
the crack faces is not zero, since the force tractions $\bt^T \bn$ and couple tractions $\bmu^T \bn$ are not vanishing as in the classical case.

\section{Crack at the interface between couple-stress elastic and classical elastic materials}
\label{sec3}

In this section we consider the problem of a crack lying at the interface between a micropolar material (occupying 
the upper half-plane) and a classical elastic one (occupying the lower half-plane). Then the governing equations 
are
\beq
\label{gov2}
\Delta w^+ - \frac{l^2}{2} \Delta^2 w^+ = 0, \quad 
\Delta w^- = 0.
\eeq
The traction-free boundary conditions along the crack faces are:
\beq
\label{bound}
p_3^+ = 0 \quad \text{and} \quad q_1^+ = 0 \quad \text{for} \quad \theta = \pi, \quad
\sigma_{23}^- = 0 \quad \text{for} \quad \theta = -\pi.
\eeq

Along the ideal interface the continuity of displacement and force tractions needs to be enforced, so that we have the 
following transmission conditions:
\beq
\label{transm}
w^+ = w^-, \quad p_3^+ = \sigma_{23}^- \quad \text{for} \quad \theta = 0.
\eeq

However, since the orders of the two governing equations are different, an additional transmission condition is 
needed. This additional condition can be chosen in two different ways. At the boundary of the micropolar material 
one can prescribe the value of the reduced couple traction $q_1^+$, or, alternatively, the value of the rotation
$\varphi_1^+$ (note that $\varphi_2^+ = \varphi_2^-$ follows immediately from (\ref{transm})$_1$). We analyse 
these two cases separately in the next subsections.

\subsection{`Couple' transmission conditions}
\label{sec31}

The additional transmission condition in this case takes the form
\beq
q_1^+ = 0 \quad \text{for} \quad \theta = 0.
\eeq

Bearing in mind that, in this case, we need to match the solutions of governing equations having different orders, 
we use here a two-terms asymptotic analysis, so that a solution is searched for in the form
\beq
\label{plus}
w^+(r,\theta) = r^\lambda F_1(\theta) + r^{\lambda+2} F_2(\theta) + O(r^{\lambda+4}), \quad r \to 0,
\eeq
\beq
\label{minus}
w^-(r,\theta) = r^\lambda H_1(\theta) + r^{\lambda+2} H_2(\theta) + O(r^{\lambda+4}), \quad r \to 0.
\eeq
Substituting (\ref{plus}) in (\ref{gov2})$_1$ and collecting like powers of $r$, we obtain
\beq
\label{system}
\Delta^2 (r^\lambda F_1) = 0, \quad
\Delta^2 (r^{\lambda+2} F_2) = \frac{2}{l^2} \Delta(r^\lambda F_1).
\eeq
For $\lambda \neq 0, 1, 2$, the system (\ref{system}) admits the solution
\beq
\label{sol1}
\barr{l}
\ds
F_1 = B_1^{(1)} \sin(\lambda\theta) + B_2^{(1)} \cos(\lambda\theta) + B_3^{(1)} \sin[(\lambda-2)\theta] + 
B_4^{(1)} \cos[(\lambda-2)\theta], \\[3mm]
\ds
\barr{l}
F_2 = B_1^{(2)} \sin[(\lambda+2)\theta] + B_2^{(2)} \cos[(\lambda+2)\theta] + B_3^{(2)} \sin(\lambda\theta) + 
B_4^{(2)} \cos(\lambda\theta) \\[1mm]
\ds + \frac{B_3^{(1)}}{4\lambda l^2} \sin[(\lambda-2)\theta] + \frac{B_4^{(1)}}{4\lambda l^2} 
\cos[(\lambda-2)\theta],
\earr
\earr
\eeq
Substituting (\ref{minus}) in (\ref{gov2})$_2$ and collecting like powers of $r$, we obtain
\beq
\label{system2}
\Delta (r^\lambda H_1) = 0, \quad 
\Delta (r^{\lambda+2} H_2) = 0.
\eeq
For $\lambda \neq 0, 1, 2$, the system (\ref{system2}) admits the solution
\beq
\label{sol2}
H_1 = A_1^{(1)} \sin \lambda\theta + A_2^{(1)} \cos \lambda\theta, \quad 
H_2 = A_1^{(2)} \sin [(\lambda+2)\theta] + A_2^{(2)} \cos [(\lambda+2)\theta].
\eeq

Substituting the two-terms asymptotics (\ref{plus}), with $F_{1,2}$ and $H_{1,2}$ given by (\ref{sol1}) and 
(\ref{sol2}) respectively, into the boundary and transmission conditions, we obtain a 12$\times$12 homogeneous algebraic 
system, whose characteristic equation is 
\beq
(1 + \eta) \cos^2(\pi \lambda) \sin^4(\pi \lambda) = 0,
\eeq
so that the exponent $\lambda$ admits the values $\lambda = k/2$, where $k = 1,3,5$. 

Note that for $\lambda = 1/2$ the first term in (\ref{minus}) for the classical elastic material corresponds to 
bounded elastic energy, while the first term in (\ref{plus}) for the micropolar material corresponds to infinite 
elastic energy. However, from the analysis of all boundary and transmission conditions, it is found that 
$F_1(\theta) \equiv 0$ in this case, so that the energetic requirements are fulfilled.

For the special cases $\lambda = 0,1,2$ the analysis is straightforward and the final asymptotic representation 
of the solution taking into account all terms $\lambda = k/2$, $k = 0,1,2,3$, is given by
$$
w^-(r,\theta) = \alpha_0 + r^{1/2} \alpha_1 \sin \frac{\theta}{2} + r \alpha_2 \cos \theta + 
r^{3/2} \alpha_3 \sin \frac{3\theta}{2} + r^2 \alpha_4 \cos 2\theta + r^{5/2} \alpha_5 \sin \frac{5\theta}{2} + 
O(r^3), \quad r \to 0, 
$$
\beq
\barr{l}
\ds w^+(r,\theta) = \alpha_0 + r (\alpha_2 \cos \theta + \beta_1 \sin \theta) + 
r^2 \left\{\alpha_4\left(\frac{1 + \eta}{2} + \frac{1 - \eta}{2}\cos 2\theta\right) + 
\beta_2 \sin 2\theta\right\} \\[3mm] 
\ds - r^{5/2} \alpha_1 \frac{2G_-}{3G_+l^2(3-\eta)} 
\left\{ \frac{3 - 5\eta}{5(1+\eta)} \sin \frac{5\theta}{2} + 
\sin \frac{\theta}{2} \right\} + O(r^3), \quad r \to 0,
\earr
\eeq
The asymptotics of stress in the lower half-plane, occupied by the classic elastic material, as $r \to 0$ are 
given by
$$
\sigma_{13}^- = -\frac{1}{2} r^{-1/2} \alpha_1 G_- \sin\frac{\theta}{2} + \alpha_2 G_- + 
\frac{3}{2} r^{1/2} \alpha_3 G_- \sin\frac{\theta}{2} + 2 r \alpha_4 G_- \cos\theta + 
\frac{5}{2} r^{3/2} \alpha_5 G_- \sin\frac{3\theta}{2} + O(r^2), 
$$
\beq
\sigma_{23}^- = \frac{1}{2} r^{-1/2} \alpha_1 G_- \cos\frac{\theta}{2} + 
\frac{3}{2} r^{1/2} \alpha_3 G_- \cos\frac{\theta}{2} - 2 r \alpha_4 G_- \sin\theta + 
\frac{5}{2} r^{3/2} \alpha_5 G_- \cos\frac{3\theta}{2} + O(r^2).
\eeq
The asymptotics of symmetric stress, couple-stress and skew-symmetric stress in the upper half-plane, occupied by 
the micropolar material, as $r \to 0$ are given by
$$
\sigma_{13}^+ = \alpha_2 G_+ + 2 r G_+ (\alpha_4 \cos\theta + \beta_2 \sin\theta) - 
\frac{2}{3} r^{3/2} \alpha_1 G_- \frac{1 - 3\eta + (5 - 3\eta)\cos\theta}{l^2(3 - \eta)(1 + \eta)} 
\sin\frac{\theta}{2} + O(r^2), 
$$
$$
\sigma_{23}^+ = \beta_1 G_+ + 2 r G_+ (\alpha_4 \eta \sin\theta + \beta_2 \cos\theta) - 
\frac{2}{3} r^{3/2} \alpha_1 G_- \frac{1 + 5\eta + (1 - 7\eta)\cos\theta}{l^2(3 - \eta)(1 + \eta)} 
\cos\frac{\theta}{2} + O(r^2), 
$$
$$
\mu_{11}^+ = 2\beta_2 G_+ l^2 (1 + \eta) - 
r^{1/2} \alpha_1 G_- \frac{1 - 3\eta + (1 + \eta)\cos\theta}{3 - \eta} \cos\frac{\theta}{2} + O(r), 
$$
\beq
\mu_{21}^+ = - r^{1/2} \alpha_1 G_- \frac{(1 + \eta)\sin\theta}{3 - \eta} \cos\frac{\theta}{2} + O(r), 
\eeq
$$
\mu_{12}^+ = -2\alpha_4 G_+ l^2 (1 - \eta^2) + 
r^{1/2} \alpha_1 G_- \frac{3 - 5\eta - (1 + \eta)\cos\theta}{3 - \eta} \sin\frac{\theta}{2} + O(r), 
$$
$$
\tau_{13}^+ = - r^{-1/2} \frac{\alpha_1 G_-}{3 - \eta} \sin\frac{\theta}{2} + O(1), 
$$
$$
\tau_{23}^+ = r^{-1/2} \frac{\alpha_1 G_-}{3 - \eta} \cos\frac{\theta}{2} + O(1).
$$
The condition (\ref{edge2}) becomes now $\mu_{22}^+(r = 0, \theta = \pi) = 0$, which requires $\beta_2$ to vanish, 
except for the case where a tangential line load is applied along the crack edge.

It is now possible to analyse the jump of the rotation component $\varphi_1$ across the interface and it is found 
that there is a mismatch between the micropolar material and the elastic one given by
\beq
\jump{0.15}{\varphi_1} = -\frac{1}{4} \alpha_1 r^{-1/2} + \frac{1}{2} \beta_1 - \frac{3}{4} \alpha_3 r^{1/2} + 
\beta_2 r + O(r^{3/2}), \quad r \to 0.
\eeq

It is concluded that in the classical elastic material the solution shows a square-root singularity, as in the 
classic case. Moreover, in the micropolar material, the symmetric stress and the couple-stress are bounded, and 
only the skew-symmetric stress shows a square-root singularity.

The energy release rate, computed in the same manner as in Sec. \ref{sec23}, is given by
\beq
J_* = \frac{\pi}{8} G_- \alpha_1^2, 
\eeq
which shows that the constant $\alpha_1$ plays the role of stress intensity factor.

\subsection{`Rotation' transmission conditions}
\label{sec32}

The problem is defined by the equations (\ref{gov2}) with the boundary and transmission conditions (\ref{bound}) 
and (\ref{transm}).
The additional transmission condition used here is the continuity of the rotation vector, namely
\beq
\varphi_1^+ = \varphi_1^- \quad \text{for} \quad \theta = 0.
\eeq
We search for a solution again in the form (\ref{plus}) and (\ref{minus}).
For $\lambda \neq 0,1,2$, the characteristic equation now takes the form
\beq
(1 + \eta)^2 \sin^2(\pi \lambda) 
[1 + \eta - (\eta - 3)\cos 2\pi \lambda]^2 = 0,
\eeq
so that the exponent $\lambda$ admits real values, namely integer positive numbers and 
\beq
\lambda = \pm \frac{1}{2\pi} \arccos \frac{\eta + 1}{\eta - 3} + k,
\eeq
where $k$ is a non negative integer.

In the particular case $\eta = 1$, the complete asymptotics of the solution up to the forth order is given by
\beq
w^-(r,\theta) = \alpha_0 + r \alpha_1 \cos \theta + r^2 \alpha_2 \cos 2\theta + 
r^3 \alpha_3 \cos 3\theta + r^4 \alpha_4 \cos 4\theta + O(r^5), \quad r \to 0,
\eeq
$$
\barr{l}
\ds w^+(r,\theta) = \alpha_0 + r \alpha_1 \cos \theta + r^2 \alpha_2 + r^{5/2} \beta_1 
\left( \cos\frac{5\theta}{2} - \cos\frac{\theta}{2} \right) + r^3 \alpha_3 
\left( \frac{3}{2} \cos \theta - \frac{1}{2} \cos 3\theta \right) \\[3mm] 
\ds + r^{7/2} \beta_2 
\left( \cos\frac{7\theta}{2} - \cos\frac{3\theta}{2} \right) + r^4 
\left\{ \left(\frac{\alpha_2}{24l^2} - \alpha_4\right) \cos 4\theta + 
\left(2\alpha_4 - \frac{\alpha_2}{6l^2}\right) \cos 2\theta + \frac{\alpha_2}{8l^2} \right\} + 
O(r^{9/2}), \quad r \to 0.
\earr
$$

The asymptotics of stress in the classic elastic material occupying the lower half-plane as $r \to 0$ become 
\beq
\barr{l}
\ds
\sigma_{13}^- = \alpha_1 G_- + 2 r \alpha_2 G_- \cos\theta + 3 r^2 \alpha_3 G_- \cos 2\theta + 
4 r^3 \alpha_4 G_- \cos 3\theta + O(r^4), \\[3mm]
\ds
\sigma_{23}^- = - 2 r \alpha_2 G_- \sin\theta - 3 r^2 \alpha_3 G_- \sin 2\theta - 
4 r^3 \alpha_4 G_- \sin 3\theta + O(r^4),
\earr
\eeq

Finally, the asymptotics of symmetric stress, couple-stress and skew-symmetric stress in the micropolar material 
occupying the upper half-plane as $r \to 0$ become
$$
\barr{l}
\ds \sigma_{13}^+ = \alpha_1 G_+ + 2 r \alpha_2 G_+ \cos\theta - 
3 r^{3/2} \beta_1 G_+ \sin\theta \sin\frac{\theta}{2} + 3 r^2 \alpha_3 G_+ \\[2mm] 
\ds - 10 r^{5/2} \beta_2 G_+ \sin^2 \frac{\theta}{2} \left(2\cos\frac{\theta}{2} + \cos\frac{3\theta}{2}\right) + 
4 r^3 \alpha_4 G_+ \cos\theta (2 - 2\cos\theta) + O(r^{7/2}),
\earr
$$
$$
\barr{l}
\ds \sigma_{23}^+ = 2 r \alpha_2 G_+ \sin\theta - 
\frac{1}{2} r^{3/2} \beta_1 G_+ \left(3\sin\frac{\theta}{2} + 7\sin\frac{3\theta}{2}\right) + 
3 r^2 \alpha_3 G_+ \sin 2\theta \\[3mm] 
\ds + \frac{1}{2} r^{5/2} \beta_2 G_+ \left(5\sin\frac{\theta}{2} - 9\sin\frac{5\theta}{2}\right) + 
\frac{2}{3 l^2} r^3 G_+ \sin\theta [\alpha_2 - (\alpha_2 - 18\alpha_4 l^2)\cos 2\theta] + O(r^{7/2}),
\earr
$$
$$
\barr{l}
\ds \mu_{11}^+ = -\frac{3}{2} r^{1/2} \beta_1 G_+ l^2 \left(5\sin\frac{\theta}{2} + \sin\frac{3\theta}{2}\right) 
+ 12 r \alpha_3 G_+ l^2 \sin\theta - 
\frac{5}{2} r^{3/2} \beta_2 G_+ l^2 \left(3\sin\frac{\theta}{2} + 7\sin\frac{3\theta}{2}\right) \\[3mm]
\ds + 24 r^2 \alpha_4 G_+ l^2 \sin 2\theta + O(r^{5/2}),
\earr
$$
\beq
\barr{l}
\ds \mu_{21}^+ = \mu_{12}^+ = 
-\frac{3}{2} r^{1/2} \beta_1 G_+ l^2 \left(5\cos\frac{\theta}{2} - \cos\frac{3\theta}{2}\right) + 
\frac{5}{2} r^{3/2} \beta_2 G_+ l^2 \left(3\cos\frac{\theta}{2} - 7\cos\frac{3\theta}{2}\right) \\[3mm]
\ds + 4 r^2 G_+ \sin^2 \theta (\alpha_2 - 12 \alpha_4 l^2) + O(r^{5/2}),
\earr
\eeq
$$
\tau_{13}^+ = \frac{3}{2} r^{-1/2} \beta_1 G_+ l^2 \cos\frac{\theta}{2} - 6 \alpha_3 G_+ l^2 + 
\frac{15}{2} r^{1/2} \beta_2 G_+ l^2 \cos\frac{\theta}{2} - 24 r \alpha_4 G_+ l^2 \cos\theta +
O(r^{3/2}),
$$
$$
\tau_{23}^+ = \frac{3}{2} r^{-1/2} \beta_1 G_+ l^2 \sin\frac{\theta}{2} - 
\frac{15}{2} r^{1/2} \beta_2 G_+ l^2 \sin\frac{\theta}{2} - 4 r G_+ (\alpha_2 - 6 \alpha_4 l^2) \sin\theta +
O(r^{3/2}).
$$
In this case, the reduced couple traction $q_1$ along the interface has the form:
\beq
q_1^+(r,\theta=0) = -6 r^{1/2} \beta_1 G_+ l^2 - 10 r^{3/2} \beta_2 G_+ l^2 + O(r^{5/2}), \quad r \to 0.
\eeq

Of course, it is possible to construct the asymptotic solution for arbitrary $|\eta| < 1$. However, the general 
form for arbitrary $\eta$ is rather complicated. As an additional example, we provide the result for 
$\eta = 1/3$ and up to the first four terms ($\lambda = 0, 1/3, 2/3, 1$):
\beq
\barr{c}
\ds
\barr{l}
\ds w^-(r,\theta) = \alpha_0 + r \alpha_1 \cos\theta + r^2 \alpha_2 \cos 2\theta + 
r^{7/3} \alpha_3 \left(\sin\frac{7\theta}{3} - \frac{1}{\sqrt{3}} \cos\frac{7\theta}{3} \right) \\
\ds + r^{8/3} \alpha_4 \left(\sin\frac{8\theta}{3} - \frac{1}{\sqrt{3}} \cos\frac{8\theta}{3} \right) + O(r^3), 
\quad r \to 0,
\earr \\[9mm]
\ds
\barr{l}
\ds w^+(r,\theta) = \alpha_0 + r \alpha_1 \cos\theta + r^2 \alpha_2 \left(\frac{2}{3} + 
\frac{1}{3}\cos 2\theta\right) \\
\ds + r^{7/3} \alpha_3 \left(\frac{10}{9}\sin\frac{7\theta}{3} - \frac{10}{3\sqrt{3}} \cos\frac{7\theta}{3} - 
\frac{7}{9} \sin\frac{\theta}{3} + \frac{7}{3\sqrt{3}}\cos\frac{\theta}{3}\right) \\
\ds + r^{8/3} \alpha_4 \left(\frac{11}{9}\sin\frac{8\theta}{3} + \frac{11}{3\sqrt{3}} \cos\frac{8\theta}{3} - 
\frac{8}{9}\sin\frac{2\theta}{3} - \frac{8}{3\sqrt{3}} \cos\frac{2\theta}{3}\right) + O(r^3),
\quad r \to 0.
\earr
\earr
\eeq

It is found that the behaviour of the displacement and stress fields is similar for any $-1 < \eta \leq 1$. In 
particular, stresses are always bounded in the classical elastic material, while in the micropolar material 
singular behaviour appears only in the skew-symmetric stress (with different level of singularity depending on 
the value of $\eta$).
Moreover, the energy release rate is always zero for any $-1 < \eta \leq 1$. This shows that this type of transmission 
conditions does not allow for the propagation of the crack along the interface, and thus has limited physical meaning.

\subsection{The special case $\eta_- = -1$ and $\eta_+ \neq -1$}
\label{sec33}
In the particular case of $\eta_- = -1$ and $\eta_+ \neq -1$, the governing equation (\ref{gov}) and the 
traction-free crack face conditions (\ref{free}) for the material in the lower half-plane are satisfied by the 
classical solution defined by the field equation
\beq
\Delta w^- = 0,
\eeq
together with the following boundary condition
\beq
\frac{\partial w^-}{\partial x_2} = 0 \quad \text{for} \quad x_1 < 0,\ x_2 = 0.
\eeq
Moreover, in this case the couple-stress and skew-symmetric stress fields in the lower half-plane identically vanish, so that 
the transmission conditions along the interface $x_1 > 0$, $x_2 = 0$ become
\beq
\label{tra2}
w^+ = w^-, \quad p_3^+ = \sigma_{23}^-,
\eeq
together with one of the following additional conditions
\beq
q_1^+ = 0, \quad \varphi_1^+ = \varphi_1^-, 
\eeq
which correspond to the cases investigated in Secs. \ref{sec31} and \ref{sec32}, respectively.

\section{Discussion and conclusions}
\label{sec4}

In the present work, the effects of strain rotation gradients on a stationary Mode III crack along the interface 
between dissimilar couple-stress elastic materials have been analytically investigated by performing an 
asymptotic analysis of the crack tip fields. It is shown that solutions without oscillations appear in the 
following two cases: when the two materials are the same (homogeneous material) and when the two materials are 
dissimilar but $\eta_+ = \eta_- = 1.$ In the latter case, the solution displays the same $r^{-3/2}$ singularity 
(appearing in the skew-symmetric stress components) as for the problem of a crack in an homogeneous CS material. 
In other cases, the solution exhibits oscillatory behaviour in the vicinity of the crack tip, with the 
overlapping zone becoming increasingly large as the ratio $\eta$ between the characteristic lengths in one of the 
materials approaches the value $-1$. The energy release rate has been calculated by means 
of the conservation $J$-integral. It is shown that contributions of the integrals along the crack faces have to be 
retained and the additional boundary condition along the crack edge (Koiter, 1964) is essential to guarantee that 
the generalized $J$-integral (Lubarda and Markenscoff, 2000) remains bounded. This additional  
boundary condition has always been omitted in the earlier literature because so far only the symmetrical problem in 
homogeneous materials was discussed. The boundary condition along the crack edge breaks the symmetry and becomes 
fundamental for interface problems. 

The special problem of a crack along the interface between couple-stress and classical elastic materials has also 
been addressed. Two types of transmission conditions have been considered: `couple' and `rotation' transmission 
conditions. In the former case, it is assumed that the couple-stress traction is continuous, and thus vanishes, 
at the interface. In the latter, it is assumed instead that the rotations are continuous at the interface. It 
turns out that the solutions are quite different in the two cases and it is not possible to satisfy 
simultaneously both type of transmission conditions, so that a mismatch is always present at the interface, 
resulting either in a non-balanced couple-stress or a discontinuity in the micro-rotations. It is shown also that 
the special case $\eta_- = -1$ and $\eta_+ \neq -1$ reduces to the problem of a crack at the interface between 
classical and couple-stress elastic materials.

\vspace{6mm}
{\bf Acknowledgements}. The paper has been completed during the Marie Curie Fellowship of A.P. at 
Aberystwyth University supported by the European Union Seventh Framework Programme under contract number 
PIEF-GA-2009-252857. E.R. gratefully aknowledges financial support from the "Cassa di Risparmio di Modena" in the 
framework of the International Research Project 2009-2010 "Modelling of crack propagation in complex materials".

%
%

\end{document}